\documentclass[12pt]{iopart}

\usepackage{dcolumn}
\usepackage{bm}
\usepackage{ifpdf}
\usepackage{hyperref}
\usepackage{dcolumn}
\usepackage{bm}
\usepackage[spanish,english]{babel}
\usepackage{amsfonts}
\usepackage{amssymb}
\usepackage{graphicx}
\usepackage[latin1]{inputenc}

\begin{document}
\title[BPS solitons in a Dirac-Born-Infeld action]{BPS solitons in a Dirac-Born-Infeld action}
\author{D. Rubiera-Garcia$^1$, C. dos Santos$^2$}
\address{$^1$ Departamento de F\'{\i}sica, Universidade Federal da Para\'{\i}ba, 58051-900 Jo\~ao Pessoa, Para\'{\i}ba, Brazil.}
\address{$^2$ Centro de F\'{\i}sica e Departamento de F\'{\i}sica e Astronomia, Faculdade de Ci\^{e}ncias da Universidade do Porto, 4169-007 Porto, Portugal.}
\ead{drubiera@fisica.ufpb.br; cssilva@fc.up.pt}
\begin{abstract}

We present several classes of solitons in ($1+1$)-dimensional models where the standard canonical kinetic term is replaced by a Dirac-Born-Infeld (DBI) one. These are static solutions with finite energy and different properties, namely, they can have compact support, or be kink or lump-like, according to the type of potential chosen, which depend on the DBI parameter $\beta$. Through a combination of numerical and analytical arguments, by which the equation of motion is seen as that corresponding to \emph{another} canonical model with a new $\beta$-dependent potential and a $\beta$-deformed energy density, we construct models in which increasing smoothly the DBI parameter both the compacton radius, the thickness of the kink and the width of the lump get modified until each soliton reaches its standard canonical shape as $\beta \rightarrow \infty$. In addition we present compacton solutions whose canonical counterparts are not compact.

\end{abstract}
\pacs{03.50.-z, 05.45.Yv, 11.27.+d}
\submitto{\JPA}
\maketitle

\section{Introduction}

Scalar fields play an important role in many areas of physics, including condensed matter physics \cite{condensed-solitons}, high energy physics through the Yukawa \cite{Rebbi} and Higgs \cite{Quigg} mechanisms, and cosmology sourcing the dark energy needed to accelerate the Universe \cite{Riess}. Within scalar fields the class of them called \emph{solitons}, namely, solutions of nondissipative field equations with finite energy and fulfilling some stability criterium (see e.g. \cite{soliton-reviews1,soliton-reviews2} for some reviews), are of special interest (note that solitons are also known to exist for gauge fields \cite{gauge-solitons}). The resemblance of some of the properties of these classical solutions with those of quantum particles has raised much interest on the study of models supporting solitons, as in the case of k-fields \cite{kfield} or in models with non-smooth V-shaped potentials \cite{Arodz0}.

K-fields are field theories where the kinetic term is non-canonical so as to avoid the constraints of Derrick's theorem \cite{Derrick}, which forbids the existence of soliton solutions in a large class of field theories. Solitons in k-field models have been intensively studied due to their applications in strong interaction physics \cite{strong}, topological defects \cite{defects,defect2} and cosmology \cite{cosmology} with the result that their properties can be quite different from the standard canonical ones. For example these solitons can have a compact support \cite{Adametal}, dubbed compactons.

Compactons are more satisfactory as classical realizations of quantum particles due to the fact that they have no infinite wings: their energy does not spread everywhere in space, instead it is concentrated in a finite region of space. As a consequence non-overlapping multi compacton configurations do not interact at all. Indeed, their only interaction arises when gravity is included. This  has been applied in the context of brane cosmology, where the compacton may act like a domain wall such that gravity confinement emerges naturally \cite{bglm} or in the dark matter context as compactons interact gravitationally without emission of particle-type radiation \cite{Adam7}.

Given the interest that solitons in k-field models have in several branches of physics, the goal of this paper is to provide several classes of soliton solutions in ($1+1$) dimensions, where the standard canonical kinetic term has been replaced by a Dirac-Born-Infeld (DBI) k-field model \cite{dbi}. DBI action was originally introduced as an attempt to remove the divergence of electron's self-energy in classical electrodynamics, and arises in the low-energy limit of string/D-Brane physics \cite{DBI-string}. In addition it has been shown to support global string solutions \cite{cosmicstring} as well as skyrmions \cite{Pavlovsky02} and their higher-dimensional counterparts, dubbed textures \cite{Ramadhan12}. We show here that it also supports other kinds of soliton solutions, namely, compactons, lumps and kinks, provided that the potential term is chosen in an appropriate way. Such solutions are obtained taking as the starting point known solitons that are supported by canonical actions, from which the noncanonical DBI counterparts deviate smoothly as the DBI parameter ($\beta$) evolves. Moreover, for any $\beta$ the noncanonical equation of motion can be seen as that corresponding to \emph{another} canonical model with a new $\beta$-dependent potential but a $\beta$-deformed energy density as compared to the DBI model, which allows to obtain insights on the behaviour of the solutions in those cases where the equation of motion cannot be analytically integrated. When $\beta \rightarrow \infty$, both the field, the potential of the noncanonical DBI model, and the $\beta$-dependent potential boil down to the standard canonical solutions.

This work is organized as follows: in Sec.\ref{sec.II} we introduce the generic formalism for k-field models and write the equation of motion for Bogomoln'yi-Prasad-Sommerfeld (BPS) states for a DBI-type action. In Sec.\ref{sec.III} we present compacton, lump and kink solutions in which their radius or thickness are changed as compared to their canonical counterparts, and in Sec.\ref{sec.IV} compact solutions whose canonical counterparts are not compact. We conclude in Sec.\ref{sec.V} with some final comments.

\section{DBI k-field model} \label{sec.II}

Let us begin by introducing the generic formalism for a k-field model in $(1+1)$ dimensions with a real scalar field $\phi$. The action is given by

\begin{equation}\label{action}
S=\int d^2 x L= \int d^2 x \left(F(X)-V(\phi)\right),
\end{equation}
where $F(X)$ is the non-canonical kinetic term, with $F$ some function of the canonical Lagrangian $X =\frac{1}{2} \partial_\mu\phi\partial^{\mu}\phi$, and $V(\phi)$ is the scalar potential. The equation of motion for the action (\ref{action}) is
\begin{equation}
\partial_{\mu} (F_X \partial^{\mu}\phi)+\partial  V/\partial \phi=0. \label{eom}
\end{equation}
where $F_X\equiv \frac{dF}{dX}$. The energy-momentum tensor

\begin{equation}
T^{\mu\nu}=F_X \partial^{\mu}\phi \partial^{\nu} \phi -\eta^{\mu\nu}L,
\end{equation}
has, for static field configurations, the non-vanishing components

\begin{eqnarray}
T^{00}&=&\varepsilon=-L=-F(X)+V(\phi) \label{energydensity}\\
T^{11}&=&-p=F(X)-2XF_X-V(\phi), \label{pressure}
\end{eqnarray}
with $\varepsilon$ and $p$ the energy density and the pressure, respectively. Note that $F(X)<0$ must be required in order to obtain a positive energy density, as a necessary condition for the associated solutions to have full physical meaning.

The equation of motion (\ref{eom}) can be integrated to get $F(X)-2XF_X-V=-c$, with $c$ an integrating constant which is identified to the pressure $p$ in Eq.(\ref{pressure}). From now on we shall deal only with BPS states \cite{Prasad}, which implies $p=0$. This is also a necessary condition for stability under Derrick's scaling, as shown in Ref.\cite{defect2}. Consequently, for BPS states the equation of motion reads

\begin{equation}\label{eq:FI}
V = F(X)-2XF_X.
\end{equation}
In this work we shall focus on the noncanonical kinetic term
\begin{equation}
F(X)=\beta^2\,\left[\sqrt{1+\frac{2X}{\beta^2}}-1\right], \label{eq:DBI}
\end{equation}
where $\beta$ is a dimensionless parameter. This Lagrangian is the natural restriction, for scalar fields, of the DBI Lagrangian \cite{dbi}, when the gauge invariant $-\frac{1}{2} F_{\mu\nu}F^{\mu\nu}=\vec{E}^2$ for electrostatic spherically symmetric fields ($\vec{E}=E(r) \vec{r}/ |r|$) is replaced by the scalar field term $X=-\frac{1}{2} \phi^{'2}$.  In the $\beta \rightarrow \infty$ limit the nonlinear modification (\ref{eq:DBI}) reduces to the canonical kinetic term, i.e., $F \sim X$.

For BPS states, the equation of motion $(\ref{eq:FI})$ for the DBI Lagrangian (\ref{eq:DBI}) becomes

\begin{equation} \label{eq:eombeta}
\frac{\phi'^2}{\beta^2}=1-\left(\frac{\beta^2}{\beta^2+ V(\phi)}\right)^2,
\end{equation}
which is a first order equation satisfying the second order equation of motion (\ref{eom}), as can be easily checked.

The energy density in (\ref{energydensity}) is given by

\begin{equation} \label{eq:energy}
\varepsilon=\frac{\phi'^2}{\sqrt{1-\frac{\phi'^2}{\beta^2}}}.
\end{equation}
This can be seen as the energy density of a canonical system corrected by a factor $(1-\frac{\phi'^2}{\beta^2})^{-1/2}$, which is similar to the relativistic mass formula for a point particle. As a consequence, the kinks, lumps and compacton solutions supported by the noncanonical DBI term (\ref{eq:DBI}) for different potentials, as introduced below, may be seen as solutions of canonical models with the very same field distribution but where the energy density profile is $\beta$-deformed, as we shall see at once. Thus they do not admit a twin-like interpretation along the lines of \cite{Twin}. Indeed, in principle, for \emph{any} model with a non-canonical kinetic term $F(X)$ and a given potential $V_{nc}(\phi)$, the field equations (\ref{eq:FI}) can always be read as those corresponding to a canonical kinetic term $X$ with a potential $V_c$, and these two potentials are related through the equation $V_{nc}=F(-V_c) + 2V_c F_{X}(-V_c)$. In addition this implies that there exists a family consisting of infinitely many members of noncanonical models having the same BPS solutions of the field equations as those of the DBI one. For the particular case we are considering here, this provides an useful tool to understand the evolution of the properties of the solitons with $\beta$, which is particularly helpful in those cases where analytical solutions cannot be obtained, since such properties can be directly read off from the potential $V_c$, thus circumventing the need to analytically solve the field equations.

Let us note that for the models considered here, as we shall see that the introduction of nonlinearities in the kinetic term is also able to change the nature of the soliton solutions as compared to their canonical counterparts ($\beta \rightarrow \infty$), from which the solution deviates as the BI parameter evolves.

\section{New soliton solutions} \label{sec.III}

\subsection{Compactons}

\begin{figure}[h]
\begin{center}
\includegraphics[width=9cm,height=5cm]{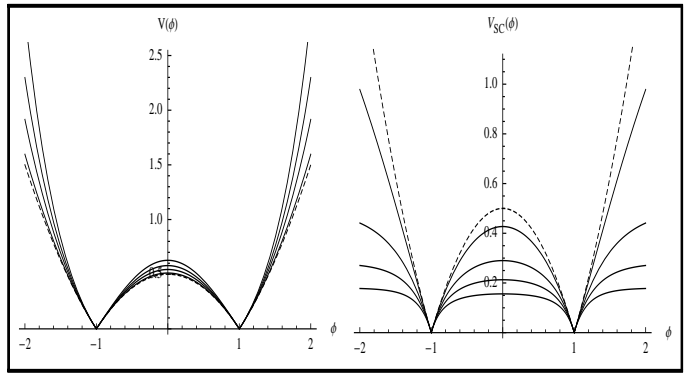}
\caption{The potentials $V(\phi)$ and $V_{SC}(\phi)$, given by Eqs.(\ref{eq:finalpotential}) and (\ref{eq:canonicalpotentialcompacton}),
respectively. Starting from the standard canonical case, namely $\beta \rightarrow \infty$ (dashed curve), which is given by Eq.(\ref{eq:standardcompactonpotential}), the set of solid curves corresponds to values $\beta=2,1,3/4,1/2$ (solid curves), and run from bottom to top in the case of $V(\phi)$, and from top to bottom for $V_{SC}(\phi)$.
\label{fig:1}}
\end{center}
\end{figure}

\begin{figure}[h]
\begin{center}
\includegraphics[width=9cm,height=5cm]{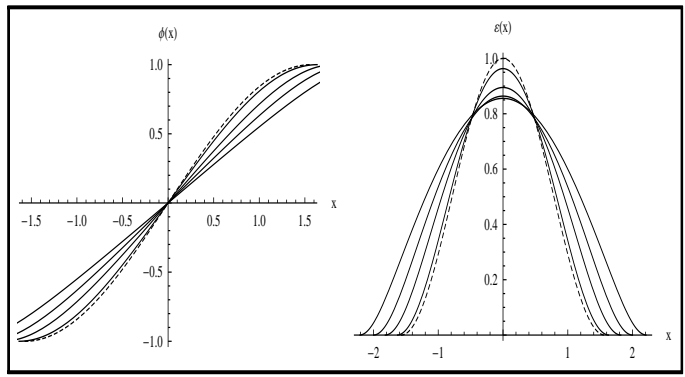}
\caption{The profile for the scalar field $\phi(x)$ and the energy density $\varepsilon(x)$ for the compacton solutions of Eq.(\ref{eq:noncanonicalcompacton}). Starting from the standard canonical case ($\beta \rightarrow \infty$, dashed curve), the set of solid curves, corresponding to $\beta=2,1,3/4,1/2$, smoothly deviate from it as $\beta$ decreases. Note that the compacton radius increases as $\beta$ decreases, while the height of the maximum of the energy density also decreases.
\label{fig:2}}
\end{center}
\end{figure}

In this section we look for compact soliton solutions, through the potential

\begin{equation} \label{eq:finalpotential}
V(\phi)=\beta^2\left(\cosh\left[\sqrt{|1-\phi^2|}/\beta\right]-1\right),
\end{equation}
which is plotted for different values of $\beta$ in Fig.\ref{fig:1}. For finite values of $ \beta$ the equation for BPS states (\ref{eq:eombeta}) becomes

\begin{equation} \label{eq:noncanonicalcompacton}
\phi'(x)=\pm \beta \tanh\left(\sqrt{|1-\phi^2|}/\beta\right).
\end{equation}
This equation has no analytical solutions and thus it has to be solved numerically. Before going that way let us note that we can interpret (\ref{eq:noncanonicalcompacton}) as the equation of motion for a scalar theory with a standard canonical kinetic term, so the associated equation of motion becomes $\phi'=\pm \sqrt{2V_{SC}}$, where ``SC" stands for ``standard canonical", and thus a potential given by

\begin{equation} \label{eq:canonicalpotentialcompacton}
V_{SC}(\phi)=\frac{\beta^2}{2}\tanh^2\left[\sqrt{|1-\phi^2|}/\beta\right],
\end{equation}
which is plotted in Fig.\ref{fig:1} for several values of $\beta$. First we note that for $\beta$ finite $V_{SC}(\phi)$ has a minima for $|\phi|=1$ while its maximum occurs at $\phi=0$. The height of this maximum, i.e., the value of the potential at $\phi=0$, decreases when $\beta$ decreases. We also note that for $\beta\rightarrow\infty$ the potential of the standard canonical case, given by

\begin{equation}\label{eq:standardcompactonpotential}
V_C(\phi) = \frac{1}{2} |1-\phi^2|,
\end{equation}
is recovered both from the potentials of Eqs.(\ref{eq:finalpotential}) and (\ref{eq:canonicalpotentialcompacton}). This brings that, on the one hand, when $\phi$ goes from $-1$ up to $+1$ it interpolates between two different minima of the potential and thus a topological defect arises. On the other hand, when $\beta$ decreases, recalling the arguments of \cite{soliton-reviews1}, due to the decrease in the well of the potential, the trajectory between the two vacua is traversed more slowly and therefore the thickness of the defect gets larger, which in this case means a larger compact radius. These analytical arguments are confirmed through the numerical plots of Fig.\ref{fig:2}, corresponding to several values of $\beta$. The solutions are compact and the compacton radius decreases as $\beta$ increases. In the $\beta \rightarrow \infty$ limit the field converges smoothly to the standard compacton solution
given by

\begin{equation} \label{eq:compactonsolution}
\phi(x)=\left\{ \begin{array}{lr} 1  & x \geq \pi/2\\
\sin(x)  &  |x|< \pi/2 \\
-1  & x \leq-\pi/2 \end{array}\right.
\end{equation}

The energy density given in (\ref{eq:energy}) is now written as
\begin{equation}
\varepsilon=\beta^2 \cosh\left(\sqrt{|1-\phi^2|}/\beta\right) \tanh^2\left(\sqrt{|1-\phi|^2}/\beta\right),
\end{equation}
and is plotted in Fig.\ref{fig:2} for several values of $\beta$. As $\beta$ decreases $\varepsilon$
spreads more in space as compared to the $\beta \rightarrow \infty$ (canonical) case, and at the same time the height of the maximum decreases. This is in agreement with the behaviour of the ``mass" term $(1-\phi^{\prime\,2}/\beta^2)^{-1/2}$ which becomes larger when $\beta$ decreases. The energy density is always positive and finite, as expected. Consequently, for the potential (\ref{eq:finalpotential}) there is a family of finite-energy compact solutions smoothly deviating from their canonical counterparts, which are recovered in the limit $\beta \rightarrow \infty$.

\subsection{Lumps}

We now proceed as in the previous section but now looking for lump-like solitons. We take the potential
\begin{equation} \label{eq:finalpotentiallump}
V(\phi)=\beta^2\left(\cosh\left[|\phi|\sqrt{1-\phi^2}/\beta\right]-1\right),
\end{equation}
which is plotted in Fig.\ref{fig:3}. Note that in the $\beta \rightarrow \infty$ limit the standard lump potential, given by

\begin{equation} \label{eq:standardlumppotential}
V_L=\frac{1}{2} \phi^2\left(1-\phi^2 \right),
\end{equation}
is recovered. In the present case the scalar field for the BPS states is the solution of the equation

\begin{equation} \label{eq:noncanonicallump}
\phi'(x)=\pm \beta \tanh\left(|\phi|\sqrt{1-\phi^2}/\beta\right).
\end{equation}
Choosing the sign $+$($-$) in (\ref{eq:noncanonicallump}), it corresponds to the region where $x$ goes
from $-\infty$(zero) up to zero ($+\infty$). The plots for $\phi(x)$ for several values of $\beta$ are in Fig.\ref{fig:4}. As expected, in the $\beta \rightarrow \infty$ limit the solutions converge smoothly to the standard lump solution given by

\begin{equation} \label{eq:standardlumpsolution}
\phi(x)=\pm sech(x).
\end{equation}

\begin{figure}[h] \label{fig:3}
\begin{center}
\includegraphics[width=9cm,height=5cm]{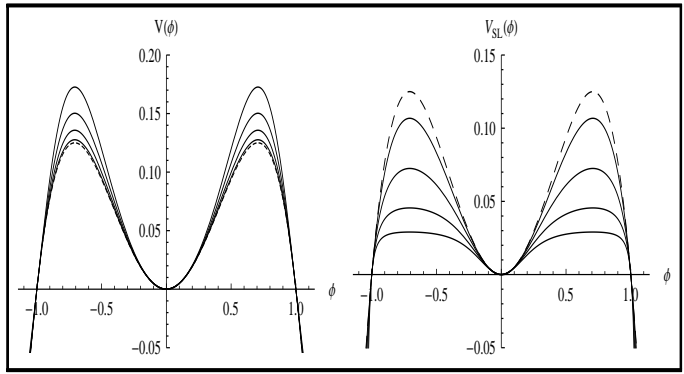}
\caption{The potentials $V(\phi)$ and $V_{SL}(\phi)$
given by Eqs.(\ref{eq:finalpotentiallump}) and (\ref{VSL}),
respectively, starting from the standard canonical case (\ref{eq:standardlumppotential}), ($\beta \rightarrow \infty$, dashed curve),  and for several values of $\beta=2,1,3/4,1/2$ (solid curves), running from bottom to top (left figure) and from top to bottom (right figure).
\label{fig:3}}
\end{center}
\end{figure}

As explained in the previous section these results can be predicted from the analysis of the potential

\begin{equation}\label{VSL}
V_{SL}(\phi)=\frac{\beta^2}{2}\tanh^2\left[|\phi|\sqrt{1-\phi^2}/\beta\right],
\end{equation}
which is obtained from seeing Eq.(\ref{eq:noncanonicallump}) as the equation of motion for a canonical model. This potential is plotted in Fig.\ref{fig:3} for several values of $\beta$.

Now, note that $V_{SL}(\phi)$ has a single minimum at $\phi=0$ for any value of $\beta$ and a maximum whose height decreases as $\beta$ decreases. It also vanishes for $\phi=0$ and for $\phi^2 = 1$ where is not a minimum. This brings that, as for the standard lump, $\phi$ evolves from zero up to zero passing through $\phi=1$ ``without stopping", i.e., $\phi$ interpolates between the same minimum of the potential and thus a non-topological defect forms. Therefore if $\phi$ travels from its zero (maximum) to its maximum (zero) one has to choose the $+(-)$ sign in Eq.(\ref{eq:noncanonicallump}).

Also note that since the height of the maximum of $V_{SL}(\phi)$ decreases when $\beta$ decreases, the lump must spread more in space as compared with the standard one. These predictions based on the behaviour of the potential
in (\ref{VSL}) are confirmed through the numerical plots of Fig.\ref{fig:4}.

\begin{figure}[h] \label{fig:4}
\begin{center}
\includegraphics[width=9cm,height=5cm]{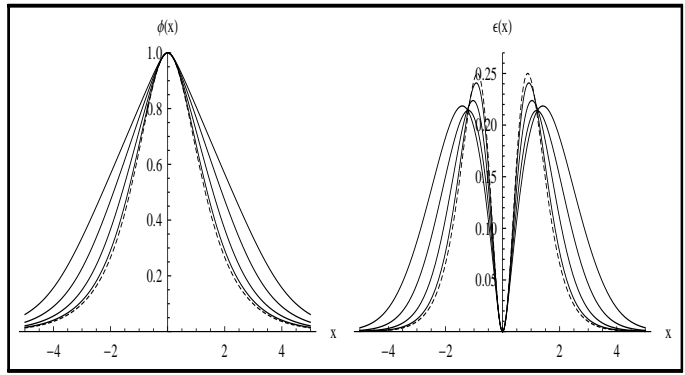}
\caption{Profile $\phi(x)$ and energy density $\varepsilon(x)$ for the lump solutions of Eq.(\ref{eq:noncanonicallump}). Starting from the standard canonical case ($\beta \rightarrow \infty$, dashed curve), the set of solid curves smoothly deviate from it and corresponds to values $\beta=1,1/2,1/3,1/4$. Note that, as $\beta$ decreases, the lump spreads more in space, and the height of the maximum of the energy density decreases. \label{fig:4}}
\end{center}
\end{figure}

The energy density is given by
\begin{equation} \label{eq:l6}
\varepsilon=\beta^2 \cosh\left(|\phi|\sqrt{1-\phi^2}/\beta \right)
\tanh^2\left(|\phi|\sqrt{1-\phi^2}\beta \right),
\end{equation}
and is plotted in Fig.\ref{fig:4} for several values of $\beta$. As $\beta$ decreases the energy density also spreads more in space while its maximum decreases. Therefore, for the potential in (\ref{eq:finalpotentiallump}) there are lump-like soliton solutions which smoothly deviate from their canonical counterpart when $\beta$ decreases.

\subsection{Kinks}

\begin{figure}[h]
\begin{center}
\includegraphics[width=9cm,height=5cm]{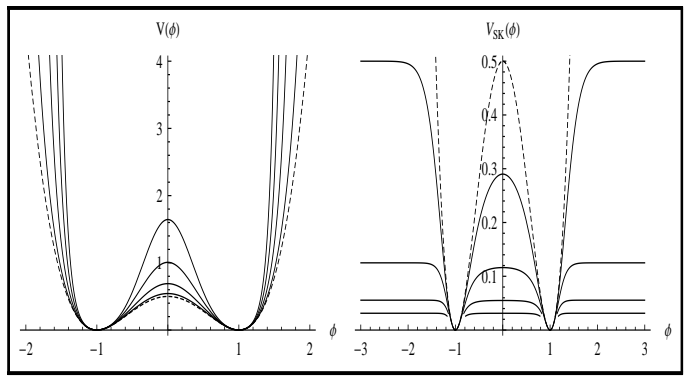}
\caption{The potentials $V(\phi)$ in Eq.(\ref{eq:finalpotentialkink}) and $V_{SK}(\phi)$ in Eq.(\ref{eq:noncanonicalkink}) for the kink solutions of Eq.(\ref{eq:k1}). The dashed curved corresponds to $\beta \rightarrow \infty$ (standard canonical case) and the solid curves correspond to $\beta=1,1/2,1/3,1/4$, running from bottom to top (left figure) and from top to bottom (right figure)}\label{fig:5}
\end{center}
\end{figure}

We now look for standard kink topological solitons (no compact support). We take the potential

\begin{equation} \label{eq:finalpotentialkink}
V(\phi)=\beta^2\left(\cosh\left[\frac{1-\phi^2}{\beta}\right]-1\right),
\end{equation}
which is plotted in Fig.\ref{fig:5} for several values of $\beta$. In the $\beta \rightarrow \infty$  limit it recovers the standard kink potential given by $V_K(\phi)=\frac{1}{2}\left(1-\phi^2 \right)^2$.

\begin{figure}[h]
\begin{center}
\includegraphics[width=9cm,height=5cm]{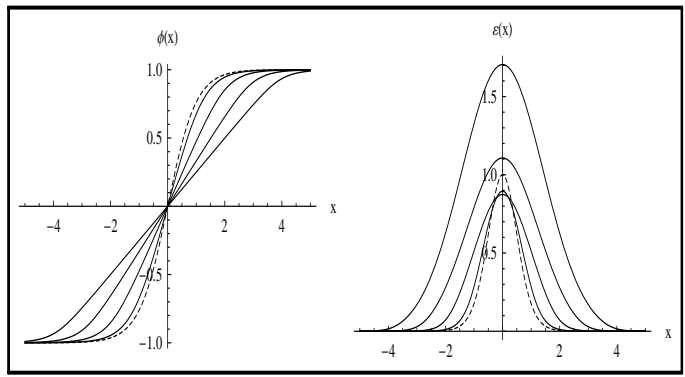}
\caption{Profile $\phi(x)$ and energy density $\varepsilon(x)$ for the kink solutions of Eq.(\ref{eq:k1}). The dashed curve is for the standard canonical case ($\beta \rightarrow \infty$), and the set of solid curves corresponds to $\beta=1,1/2,1/3,1/4$. Note that, as discussed in the text, the thickness of the kink increases as $\beta$ decreases.}\label{fig:6}
\end{center}
\end{figure}

For $\beta$ finite the equation of motion reads
\begin{equation} \label{eq:k1}
\phi'(x)=\pm \beta \tanh\left(\frac{1-\phi^2}{\beta}\right),
\end{equation}
which again does not admit analytical solutions and thus we have to solve it numerically. In Fig.\ref{fig:6} we plot the behaviour of the kink profile as a function of $\beta$. We note that these solutions smoothly deviate from their standard canonical counterpart $\phi(x)=\tanh(x)$,
giving rise to kink solutions which are more spread in space. The $\beta$-dependent ``canonical" potential to be analyzed is now
\begin{equation} \label{eq:noncanonicalkink}
V_{SK}(\phi)=\frac{\beta^2}{2}\tanh^2\left[\frac{1-\phi^2}{\beta}\right],
\end{equation}
which is plotted in Fig.\ref{fig:5} for several values of $\beta$. It has two minima at $|\phi|=1$ for any $\beta$ value and a maximum whose height decreases as $\beta$ decreases. As a result $\phi$ interpolates between the vacua of the potential and thus a topological kink defect forms. When $\beta$ decreases the kink travels more slowly between such vacua, which means that its ``thickness" increases, as confirmed directly in Fig.\ref{fig:6}. On the other hand, the energy density is given by

\begin{equation} \label{eq:l6}
\varepsilon=\beta^2 \cosh\left(\frac{1}{\beta}(1-\phi^2) \right)
\tanh^2\left(\frac{1}{\beta}(1-\phi^2) \right),
\end{equation}
and is plotted in Fig.\ref{fig:6}, where it is shown that it is always positive for any $\beta$. As $\beta$ decreases
the energy density becomes more spread in space while its maximum decreases. Therefore our model with the potential in (\ref{eq:finalpotentialkink}) supports kink-like soliton solutions which smoothly deviate from their canonical counterpart as $\beta$ decreases.

\section{Compactons without compact canonical counterparts} \label{sec.IV}

\begin{figure}[h]
\begin{center}
\includegraphics[width=9cm,height=5cm]{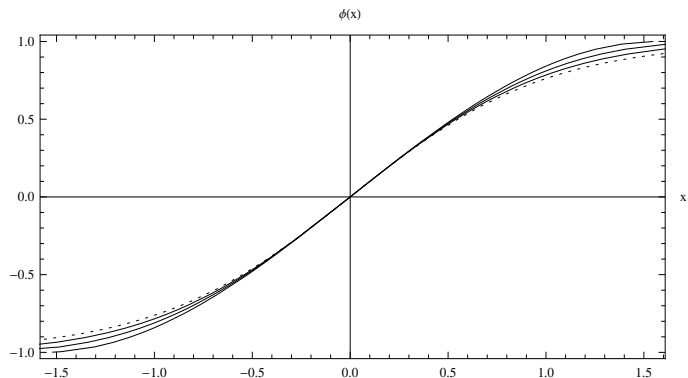}
\caption{The scalar field (\ref{solutionkinkc}) for $\beta=2,5,10,\infty$ (dashed).
\label{fig:8}}
\end{center}
\end{figure}

Let us now consider two models where the effect of the DBI action is to change not only the thickness of the standard soliton solution but also its nature. For the first model take the potential
\begin{equation}\label{potentialkinkc}
V(\phi) =\beta^2\,\left[-1+\left(1-\left(1-\phi^2\right)^{\frac{2\beta}{1+\beta}}/\beta^2\right)^{-1/2}\right].
\end{equation}
The equation for BPS states $(\ref{eq:eombeta})$ reads now

\begin{equation} \label{eq:eombetakinkc}
\phi^\prime=\pm\left(1-\phi^2\right)^{\frac{\beta}{1+\beta}},
\end{equation}
and the respective energy density in (\ref{eq:energy}) is given by

\begin{equation} \label{eq:energykinkc}
\varepsilon=\left(1-\phi^2\right)^{\frac{2\beta}{1+\beta}}
\left[1-\left(1-\phi^2\right)^{\frac{2\beta}{1+\beta}}/\beta^2\right]^{-1/2}.
\end{equation}
Let us note that in order to have a definite positive energy density at $\phi \rightarrow 0$ the $\beta$ parameter defining the model must satisfy $\beta>1$. The solutions to (\ref{eq:eombetakinkc}) are analytical and given, in implicit form, by

\begin{equation} \label{solutionkinkc}
x=\pm \phi \cdot 2F1 \left[\frac{1}{2},\frac{\beta}{\beta+1},\frac{3}{2},\phi^2\right],
\end{equation}
where $2F1$ is an hypergeometric function. These solutions are plotted in Fig.\ref{fig:8} for several values of $\beta$. They correspond to compact solitons whose shape is kink-like and whose compacton radius can be exactly computed as

\begin{equation} \label{radiuskinkc1}
R =\frac{\sqrt{\pi}}{2}\frac{\Gamma\left(\frac{1}{1+\beta}\right)}{\Gamma\left(\frac{3+\beta}{2+2\beta}\right)},
\end{equation}
where $\Gamma$ is the Euler Gamma function. This radius grows almost linearly from $1$ to $\infty$, while the solution varies smoothly (see Fig. (\ref{fig:8})) between the compact one (\ref{eq:compactonsolution}) with radius
$\pi/2$ and the standard kink solution $\phi=\tanh(x)$
which has no compact support ($R\rightarrow \infty$ when $\beta\rightarrow \infty$ in Eq.(\ref{radiuskinkc1})).

\begin{figure}[h]
\begin{center}
\includegraphics[width=9cm,height=5cm]{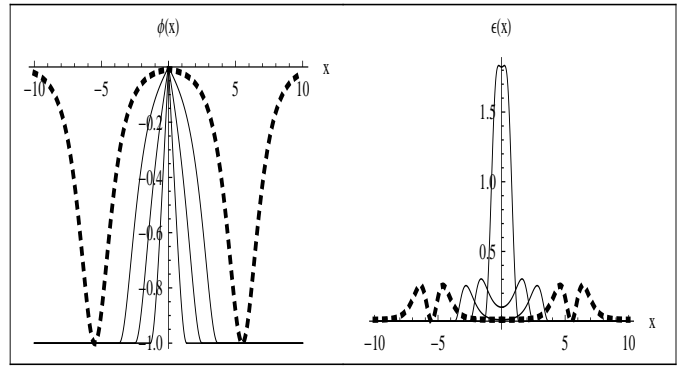}
\caption{The scalar field (\ref{eq:compactonsolution}) and the energy density (\ref{eq:energylumpc}) for several values $\beta=1.1,10,100$, as compared to the canonical system ($\beta \rightarrow \infty$, dashed).
\label{fig:12}}
\end{center}
\end{figure}

As the second model, let us consider the potential

\begin{equation}\label{potentiallumpc}
V(\phi) =\beta^2\,\left[\left(1-(\phi^2+\frac{1}{\beta})(1-\phi^2)/\beta^2\right)^{-1/2}-1\right].
\end{equation}
The BPS solutions to the equation of motion

\begin{equation} \label{eq:eombetalumpc}
\phi^\prime=\pm\left((\phi^2+\frac{1}{\beta})(1-\phi^2)\right)^{1/2},
\end{equation}
are analytic and given by (see Fig.\ref{fig:12})

\begin{equation} \label{eq:compactonsolution}
\phi(x)=\left\{ \begin{array}{lr} JacSN\left[\frac{-Q+x}{\sqrt{\beta}},-\beta\right];  & Q-A<x<Q\\
JacSN\left[\frac{Q-x}{\sqrt{\beta}},-\beta\right];   & Q < x < Q+A \end{array}\right.
\end{equation}
where $JacSN\left[u,m\right]$ is the Jacobi elliptic function. These solutions correspond to compact solitons whose shape is lump-like centered at $Q$ and whose radius is given by
\begin{equation} \label{radiuslumpc1}
R=\sqrt{\beta} EllipticK\left[-\beta\right],
\end{equation}
increasing as $\beta$ decreases from $\beta \rightarrow \infty$, where $R \rightarrow \infty$, which means that the canonical solutions are not compact. Without loss of generality we can take $Q=0$ for which the energy density in (\ref{eq:energy}) is given by

\begin{equation} \label{eq:energylumpc}
\varepsilon=\left(1-\phi^2\right)\left(\frac{1}{\beta}+\phi^2\right)
\left[1-\left(1-\phi^2\right)\left(\frac{1}{\beta}+\phi^2\right)/\beta^2\right]^{-1/2},
\end{equation}
which is plotted in Fig.\ref{fig:12} for several values of $\beta$. Again, for positive definiteness of the energy density at $\phi \rightarrow 0$, the constraint $\beta>1$ must be satisfied.

\section{Final comments} \label{sec.V}

In this letter we have presented models including a DBI noncanonical kinetic term and supporting ($1+1$)-dimensional static solitons of compacton, lump and kink type, depending on the potential chosen. Through a combination of numerical and analytical arguments using another canonical model with a new $\beta$-dependent potential and a $\beta$-deformed energy density, we have shown that by varying $\beta$ either the compacton radius, the thickness of the kink, or the width of the lump are modified until each soliton reaches its standard canonical counterpart shape in a smooth way. Moreover we have constructed two families of compacton solutions whose canonical counterparts are not compact. This implies that k-field models may provide modifications on the properties and nature of the canonical soliton solutions. Since solitons are of direct interest to several branches of physics, such as solid state physics, cosmology or strong-interaction models, our study of DBI k-fields supporting such solution solutions may provide useful tools for the analysis of some of these systems.

\section*{Acknowledgments}

D. R.-G. is funded by CNPq (Brazilian agency) through grant number 561069/2010-7, and thanks the Departamento de F\'{\i}sica e Astronomia da Faculdade de Ciencias da Universidade do Porto for their hospitality.\\

\end{document}